\documentclass[a4paper,11pt]{article}
\usepackage{pos}
\usepackage{subcaption}
\title{Investigating peculiar prompt emission properties of the multi-Peaked GRB~250129A}

\author*[a]{Ankur Ghosh}
\author[a,b]{Tamador K.~M.~Aldowma}
\author[a,c,d]{Soebur Razzaque}
\affiliation[a]{Centre for Astro-Particle Physics (CAPP) and Department of Physics, University of Johannesburg, PO Box 524, Auckland Park 2006, South Africa}

\affiliation[b]{Department of Astronomy and Meteorology, Faculty of Science and Technology, Omdurman Islamic University, PO Box 382, Omdurman, 14415, Sudan}

\affiliation[c]{Department of Physics, The George Washington University, Washington, DC 20052, USA}

\affiliation[c]{National Institute for Theoretical and Computational Sciences (NITheCS), Private Bag X1, Matieland, South Africa}

\emailAdd{ghosh.ankur1994@gmail.com}
\emailAdd{srazzaque@uj.ac.za}

\abstract{We present a high-energy spectral analysis of GRB~250129A, which was triggered by the \textit{Swift}-BAT. The burst exhibits a complex, multi-peaked temporal structure characterized by two distinct emission episodes, with the main peak occurring approximately 180 seconds after the BAT trigger. The time-integrated spectral analysis in the 15–150 keV energy range indicates that a broken power-law (BPL) model provides the best fit, signifying a non-thermal origin of the prompt emission. A time-resolved spectral analysis, performed using the Bayesian block technique, shows that the intervals around the main emission peak are well described by the BPL model, while the fits for low-count intervals remain less constrained. An evident intensity-tracking behaviour is observed between the flux and the spectral peak energy ($E_p$). Furthermore, both the Amati relation and hardness–intensity correlation suggest that GRB~250129A occupies an intermediate regime, acting as a bridge between long and ultra-long GRBs.

}

\FullConference{High Energy Astrophysics in Southern Africa (HEASA2025)\\
16-20 September, 2025\\
University of Johannesburg, South Africa\\}

\begin{document}
\maketitle

\section{Introduction}
\label{sec1:intro}

Gamma-ray bursts (GRBs) are brief, extremely luminous explosions of $\gamma$-rays that release tremendous energy ($10^{48}$–$10^{55}$ erg) within just a few seconds, allowing for their detection from cosmological distances \cite{2015PhR...561....1K}. Based on the duration of their prompt emission ($T_{90}$), GRBs are generally divided into two main categories: long/soft and short/hard bursts \cite{1993ApJ...413L.101K}. Long-duration GRBs are typically linked to the collapse of massive stars, leading to the birth of a black hole \cite{2006ARA&A..44..507W}, whereas short-duration bursts are associated with the mergers of compact binaries such as neutron star–neutron star or neutron star–black hole systems \cite{2017PhRvL.119p1101A}. Another subclass of GRBs, know as the ultra-long GRBs, are characterised by the longer prompt emission lasting \(\gtrsim 10^{3}\)–\(10^{4}\) s \citep{2014ApJ...781...13L}. Their timescales have been connected to the progenitor systems such as blue supergiant collapses or tidal disruption events \citep{2014ApJ...781...13L}.

Even after more than five decades of study, the exact nature of the emission mechanisms and the locations of the radiating regions in GRBs remain unresolved. According to the fireball model \cite{1999PhR...314..575P}, the prompt emission originates from two components: a thermal photospheric emission that emerges when the jet becomes optically thin, and a non-thermal component resulting from internal shocks or magnetic reconnection within the relativistic outflow. Although this model predicts a prominent thermal contribution, 
observational evidence suggests that most GRB spectra are predominantly non-thermal. They are typically characterized by empirical models such as the Band function \cite{1993ApJ...413..281B}, cutoff power-law (\texttt{CPL}), or broken power-law (\texttt{BPL}), either individually or in combination. The prompt emission phase frequently shows complex, multi-peaked temporal profiles, which imply rapid fluctuations in the activity of the central engine \cite{2006ApJ...642..354Z}. Each distinct pulse in the light curve corresponds to a separate episode of energy release, often attributed to intermittent accretion or magnetically driven instabilities near the compact object. Understanding the causes of these temporal variations provides valuable insight into the processes governing jet dynamics, energy dissipation, and radiation mechanisms in GRBs \cite{2011ApJ...726...90Z}.

GRB~250129A ($z = 2.151$ \cite{2025GCN.39071....1S}), detected by the \textit{Swift} Burst Alert Telescope (BAT), exhibits a complex multi-peaked structure and a total $T_{90}$ duration of approximately $\sim260$ s, placing it in the category of long GRBs. In this work, we perform both time-integrated and time-resolved spectral analyses to examine the emission properties and radiation mechanisms of this burst. 

The paper is structured as follows: Section~\ref{ch2:data} outlines the BAT data reduction and analysis methods; Section~\ref{ch3:results} presents the spectral modeling results and their comparison with other GRBs; and Section~\ref{ch4:summary} summarizes the key conclusions. Throughout this study, we adopt a flat $\Lambda$CDM cosmology with parameters $H_0 = 69.6~\rm km~s^{-1}~Mpc^{-1}$, $\Omega_m = 0.286$, and $\Omega_\Lambda = 0.714$ \cite{2014ApJ...794..135B}.

\section{\textit{Swift}-BAT data reduction and analysis}
\label{ch2:data}

\begin{figure}
    \centering
	\includegraphics[width=0.5\columnwidth]{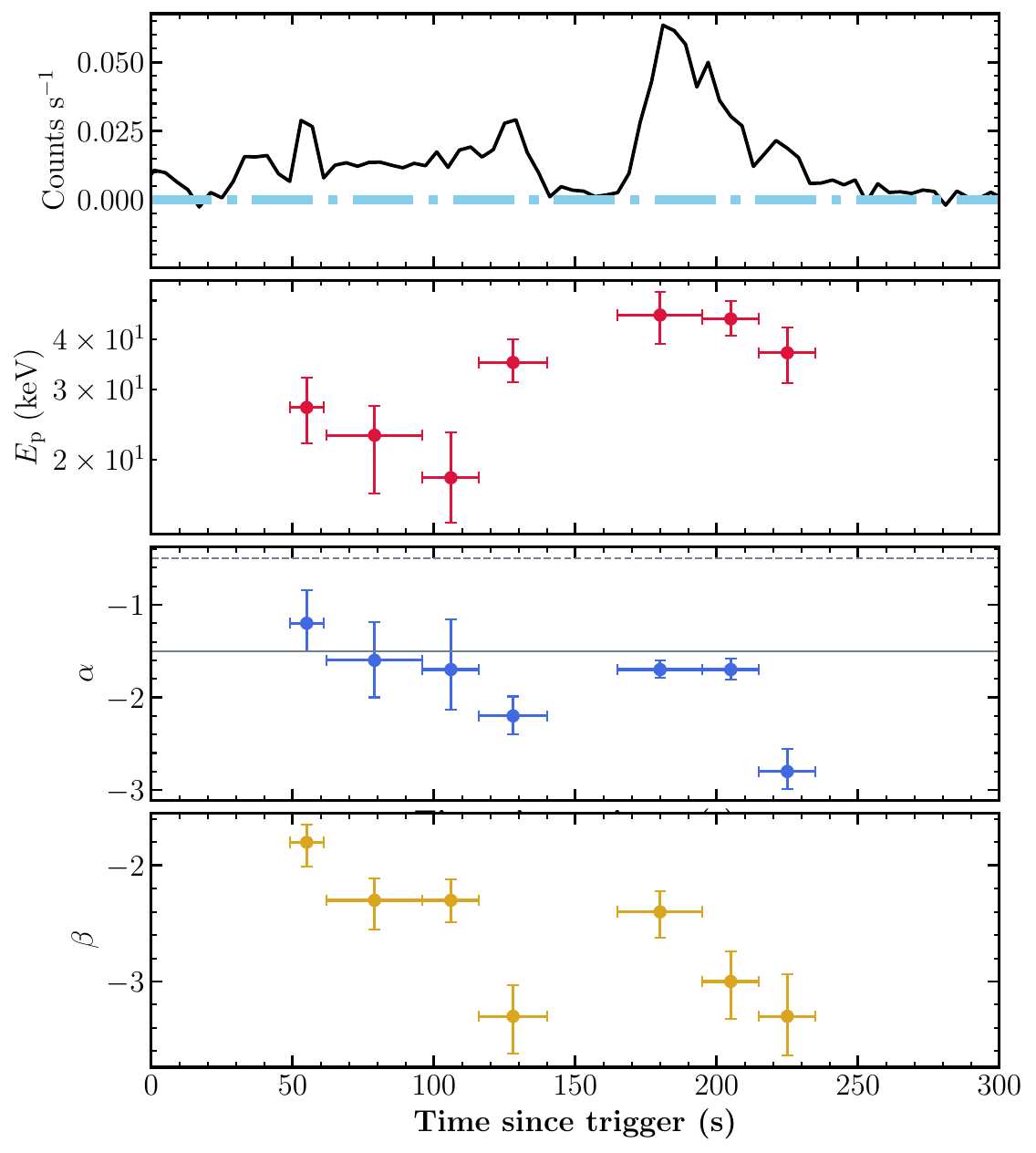}
    \caption{Evolution of spectral parameters obtained for GRB 250129A from the time-resolved spectral analysis: (first row) The background subtracted mask weighted light curve with time binning of 4 s where the blue dash dotted line represents the background level, (second row) The evolution of the peak energy (the crimson circles), (third row) The evolution of the low-energy spectral index (the indigo circles) using the \textit{Swift}-BAT data. Grey solid and dashed line indicate lines of death for synchrotron fast cooling ($\alpha$ = $-3/2$) and slow cooling ($\alpha$ = $-2/3$), respectively, and (fourth row) the evolution of high-energy spectral index.}  
    \label{fig:time_reolved}
\end{figure}

\textit{Swift}–BAT \cite{2005SSRv..120..143B} data of GRB~250129A was retrieved from the \textit{Swift} data archive portal hosted by the UK \textit{Swift} Science Data Centre\footnote{\url{https://www.swift.ac.uk/swift_portal/}}. The reduction and analysis were performed using the latest version of the \texttt{HEASOFT} software package (v6.36). The detector plane image (DPI) was generated, and hot pixels were identified and corrected using the tasks \texttt{batbinevt} and \texttt{bathotpix}, respectively. Background subtraction on the event data was carried out with the \texttt{batmaskwtevt} pipeline to obtain clean, mask-weighted events. Subsequently, the \texttt{batbinevt} tool was used to extract the light curves, which are presented in the first row of Fig.~\ref{fig:time_reolved}. The \textit{Swift}-BAT data reveals a complex multi-peaked temporal profile with two distinct emission episodes. The $T_{90}$ (15-350 keV) duration of GRB~250129A, as calculated for BAT (see Table \ref{fig:time_reolved}), is 262.25 $\pm$ 23.71 s \cite{2025GCN.39147....1M}. The first episode is relatively weak, with a duration of $\sim$ 140 s from the trigger. This episode does not show any single prominent peak. In contrast, the second episode lasts for $\sim$ 100  seconds (162 - 262 s post trigger) with the peak at $\sim \ T_0$ + 180 s.

\begin{table}
\caption{Burst properties of GRB~250129A}  
\label{200613A_prop}
\vspace{-0.6cm}
\begin{center}
\begin{tabular}{ |c||c||c| } 
\hline
Parameters & Values & References \\
\hline\hline
$T_{90}$ duration (s) &  $262.25 \pm 23.71$ & \citet{2025GCN.39147....1M}\\
Redshift & 2.151 & \citet{2025GCN.39071....1S}\\
Fluence (erg cm$^{-2}$)  &   ($5.0 \pm 0.2) \times 10^{-6}$ & \citet{2025GCN.39147....1M} \\
$E_{\rm iso}$ (erg) & $(5.48 \pm 0.22) \times 10^{52}$ & This work\\
$E_{\rm p}$ (keV) &  $38 \pm 0.22$  	& This work\\
\hline
\end{tabular}
\end{center}
\vspace{-0.5cm}
\end{table}

\begin{figure}
     \centering
     \begin{subfigure}[b]{0.50\textwidth}
         \centering
         \includegraphics[width=\linewidth]{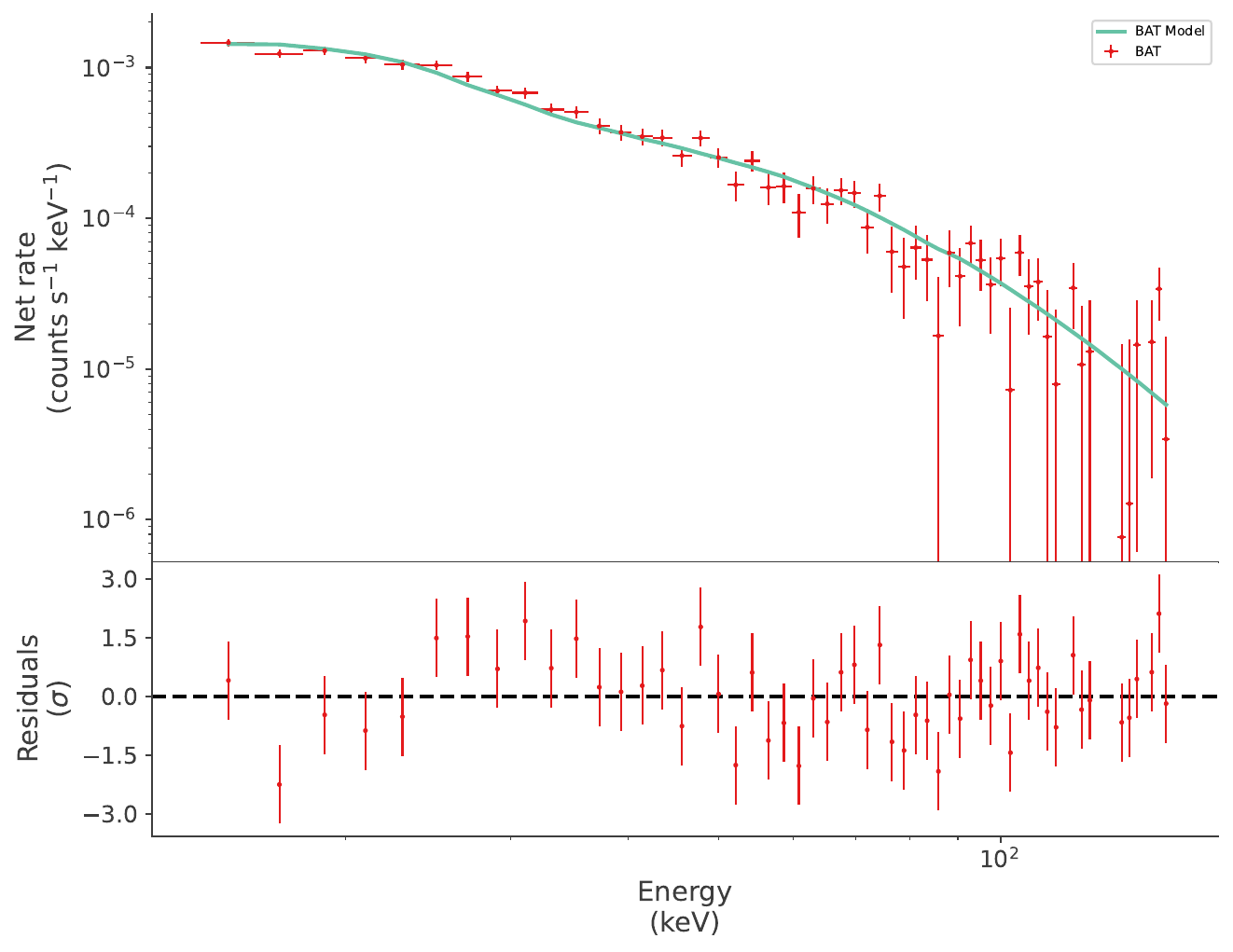}
         \caption{}
         \label{fig:lc}
     \end{subfigure}
     \hfill
     \begin{subfigure}[b]{0.45\textwidth}
         \centering
         \includegraphics[width=\linewidth]{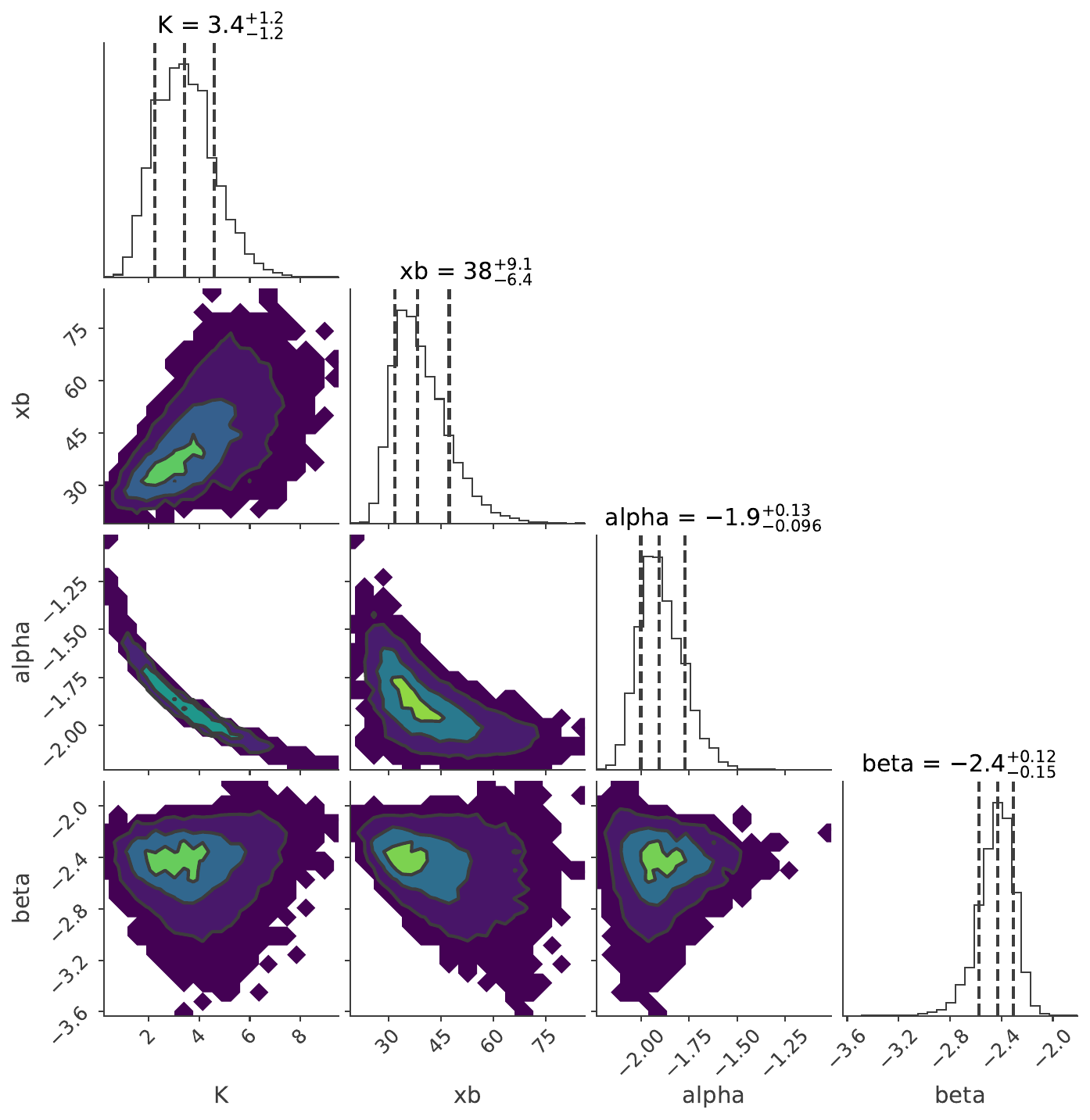}
         \caption{}
         \label{fig:corner}
     \end{subfigure}
     \hfill
        \caption{Time-integrated \textit{Swift}/BAT spectrum (15-150 keV) of GRB~250129A was best fitted with a broken power-law model. The bottom panel presents the residuals between the data and the model. (b) Posterior distributions of the broken power-law model parameters obtained from the fitting and optimisation using dynasty-nested.} 
        \label{fig:spec}
\end{figure}

The \texttt{batbinevt} task was employed to generate the Pulse Height Analyzer (PHA) file from the event data, which was subsequently cleaned to remove hot pixels and mask-weighted for background correction. The PHA was corrected for the geometrical and systematic errors using \texttt{batphasyserr} and \texttt{batupdatephakw} tools. In order to model the BAT spectrum, we generated a detector response matrix using tool called \texttt{batdrmgen}. Due to the unavailability of prompt emission data by other satellites, the joint spectral fit was not possible. The time-integrated BAT spectrum from 15-150 keV of GRB~250129A was reduced with the Multi Mission Maximum Likelihood framework (3ML\footnote{\url{https://threeml.readthedocs.io/en/latest/}}, \cite{2015ICRC...34.1042V}). Multiple phenomenological models such as the \texttt{Band} function \cite{1993ApJ...413..281B}, \texttt{CPL}, \texttt{BPL}, black body (\texttt{BB}), as well as their combinations were used to fit the time-integrated spectra. Bayesian parameter estimation was performed using the built-in ‘dynasty-nested’ sampler in 3ML, and the best-fit model was determined based on the Bayesian Information Criterion (BIC) and log-evidence values (see Fig.~\ref{fig:corner}). The time-integrated spectrum of GRB~250129A is best described by a BPL model, with the corresponding best-fit parameters summarized in Table~\ref{tab:BAT_spec}.

\begin{table}
\caption{Model parameters from the time-integrated spectral analysis of GRB~250129A for the full \textit{Swift}-BAT interval (0--262 s).}
\label{tab:BAT_spec}
\vspace{-0.6cm}  
\begin{center}
\setlength{\tabcolsep}{3pt} 
\renewcommand{\arraystretch}{1.1} 
\begin{tabular}{|p{2.0cm}|p{2.0cm}||p{2.0cm}|p{2.0cm}||p{2.0cm}|p{2.0cm}|}
\hline
BPL & Best fit value & CPL & Best fit value & Band & Best fit value \\
parameters & & parameters & & parameters & \\
\hline
$\alpha$ & $-1.90_{-0.09}^{+0.13}$ & index & $-1.5_{-0.06}^{+0.06}$ & $\alpha$ & $-1.20_{-0.28}^{+0.27}$ \\
$E_b$ (keV)$^{1}$ & $38.0_{-6.4}^{+9.1}$ & $E_c$ (keV)$^{2}$ & $62.00_{-4.90}^{+4.60}$ & $E_p$ (keV)$^{3}$ & $30.0_{-2.7}^{+2.6}$ \\
$\beta$ & $-2.40_{-0.15}^{+0.12}$ & & & $\beta$ & $-2.40_{-0.17}^{+0.11}$ \\
BIC & $84.11$ & BIC & $88.66$ & BIC & $92.72$ \\
$\log(Z)^{4}$ & $-16.92$ & log(Z) & $-16.67$ & log(Z) & $-15.13$ \\
\hline
\end{tabular}
\end{center}
\footnotesize{$^{1}$ Break energy; $^{2}$ Cutoff energy; $^{3}$ Peak energy; $^{4}$ Evidence}
\vspace{-0.2cm} 
\end{table}

\subsection{Time resolved spectral analysis}
\label{time-resolved}

Time-resolved spectral analysis is a crucial tool to inspect the pulse-wise radiation mechanism of GRB prompt emission and to study the correlations among the spectral parameters which is still not fully understood. We have sliced the BAT light curve into 12 time intervals with sufficient counts based on the Bayesian block algorithm to generate time-resolved spectra of each interval. The same set of spectral models used for the time-integrated analysis was also applied to the time-resolved spectra. Based on the lowest log-evidence values, the broken power-law (BPL) model is selected as the best fit. All three intervals within the second emission episode of GRB~250129A (160–170 s post-trigger) are well described by the BPL model, whereas only four out of the eight time intervals with higher count-rate in the first episode favour the BPL fit. Rest of the intervals are not fitting well with the model due poor count statistics which should be modelled more carefully. The evolution of the BPL model parameters are shown in Fig.~\ref{fig:time_reolved}. 

The evolution of $E_b$ closely follows the count rate variation, displaying a distinct intensity-tracking behavior throughout the entire emission phase of the burst. Likewise, the evolution of  $\alpha$ does not show a similar trend during the first emission episode. However, $\alpha$ follows the decaying trend during the second episode. $\alpha$ often exceeds the synchrotron fast-cooling limit \cite{2013ApJS..208...21G} for a few bins, but always remains under the slow cooling limit which clearly indicates the synchrotron origin of the prompt emission.

\section{GRB correlations}
\label{ch3:results}

The spectral parameters derived from both the time-integrated and time-resolved analyses of GRB~250129A can be compared with representative samples of long and short GRBs to investigate its physical nature and possible classification.

\subsection{Amati correlation}
\label{Amati}

\begin{figure}
     \centering
     \begin{subfigure}[b]{0.48\textwidth}
         \centering
         \includegraphics[width=\linewidth]{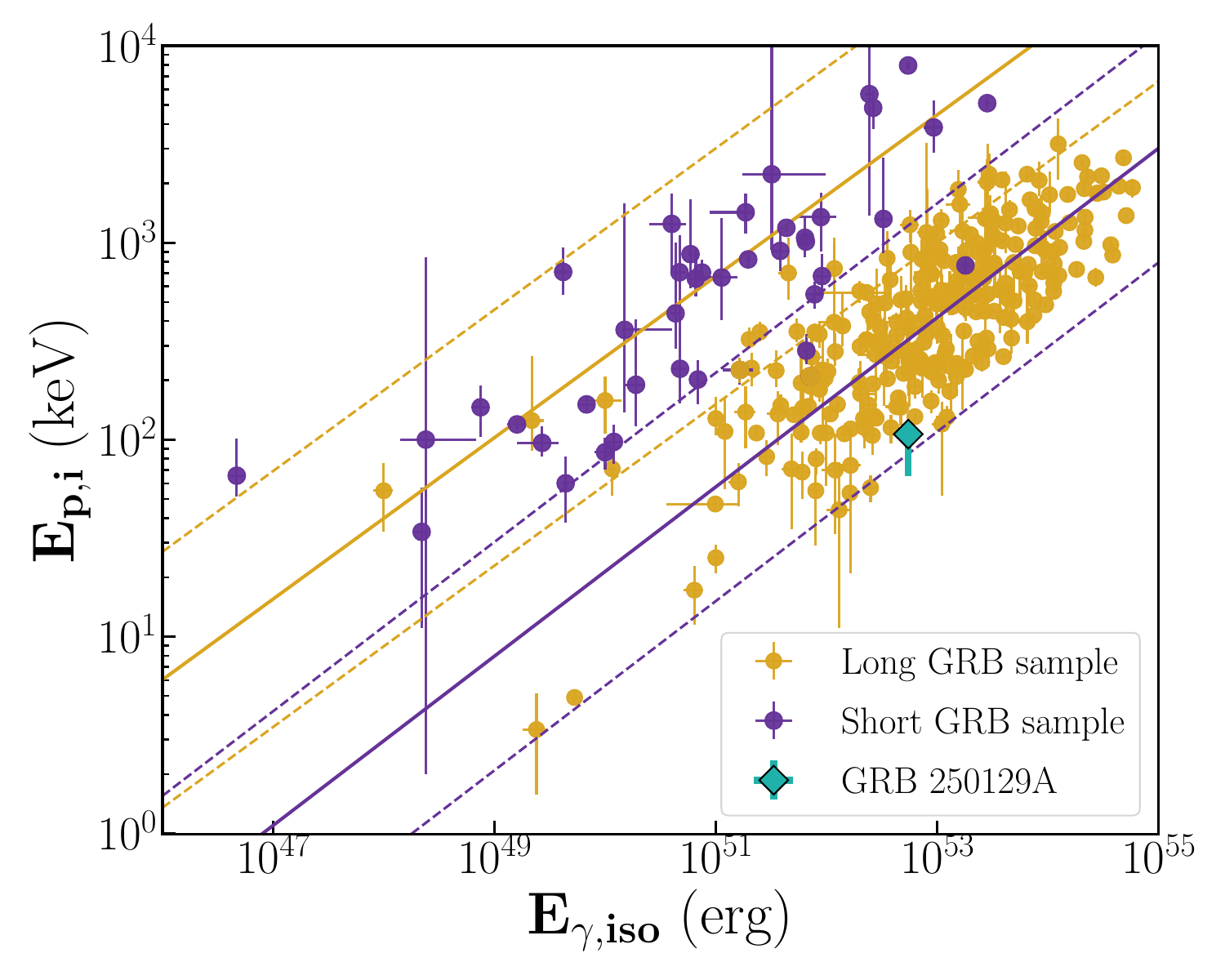}
         \caption{$E_{\rm p,i}$--$E_{\rm iso}$ correlation}
         \label{fig:amati}
     \end{subfigure}
     \hfill
     \begin{subfigure}[b]{0.48\textwidth}
         \centering
         \includegraphics[width=\linewidth]{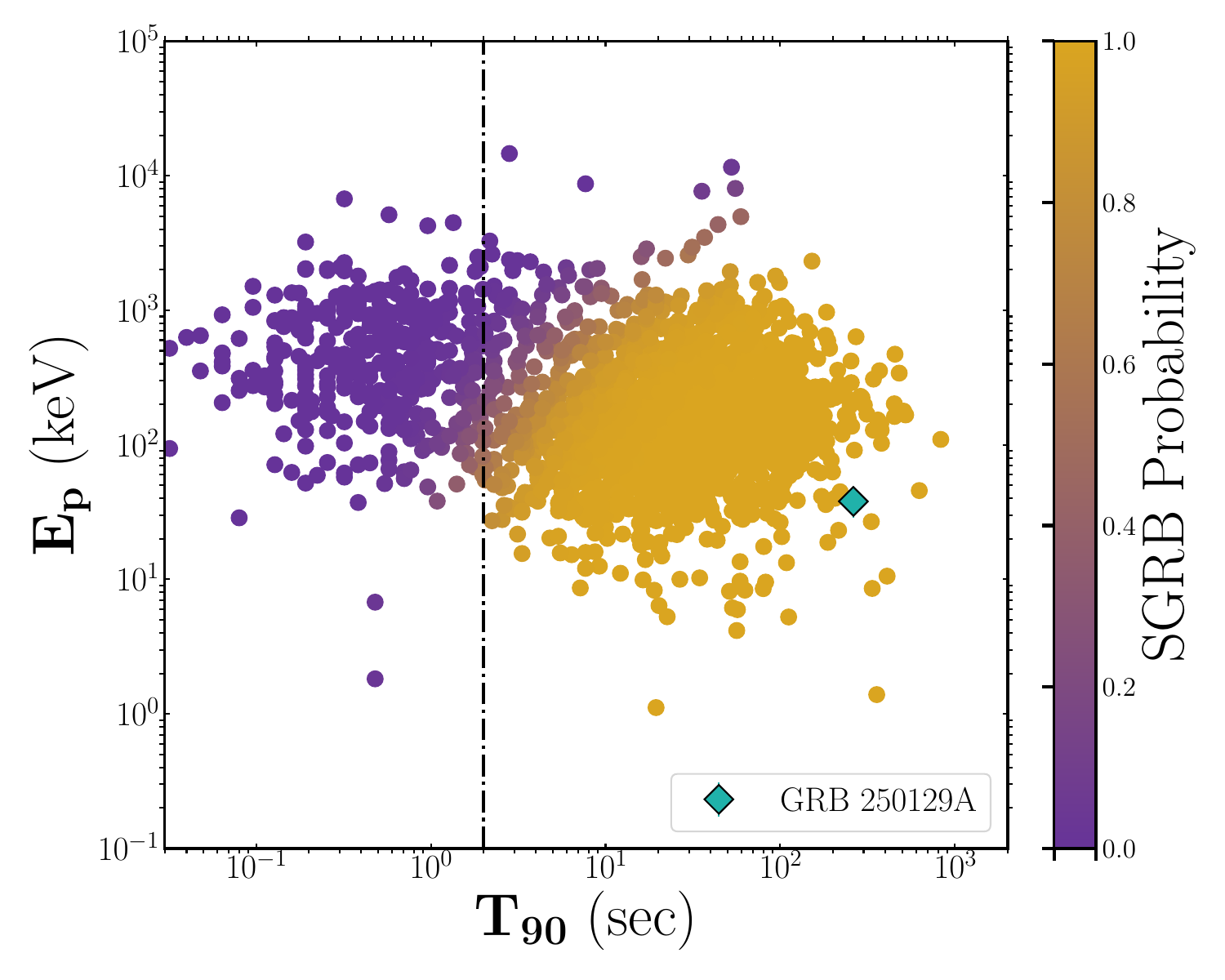}
         \caption{$E_{\rm p} - T_{90}$ correlation}
         \label{fig:hardness}
     \end{subfigure}
     \hfill
        \caption{The $E_{\rm p,i}$--$E_{\rm iso}$ and $E_{\rm p}$--$T_{90}$ correlation for the sample of long (golden circles) and short (purple circles) GRBs. In the $E_{\rm p,i}$--$E_{\rm iso}$ plot, the dashed line in the same color corresponds to the 3-$\sigma$ scatter of the correlation. The light-green diamond symbol represents the position of GRB~250129A in these plots. Black dashed line in Fig. 3(b) indicates the boundary between short and long GRBs at $T_{\rm 90}$ = 2 s.} 
        \label{fig:corr}
\end{figure}

A well-established diagnostic for distinguishing between short and long GRBs is the correlation between the intrinsic spectral peak energy of the $\nu F_{\nu}$ spectrum ($E_{\rm p,i}$) and the isotropic-equivalent energy released during the prompt emission phase - referred to as the Amati relation \cite{2006MNRAS.372..233A}. Figure~\ref{fig:amati} illustrates this relation using a comparative sample of 316 GRBs compiled by \cite{2021MNRAS.504..926M}. The isotropic-equivalent gamma-ray energy ($E_{\gamma,\mathrm{iso}}$) for GRB~250129A was computed using the BAT-measured fluence in the 15–150~keV range, $S_{\gamma} = (5.0 \pm 0.2) \times 10^{-6}\,\,\mathrm{erg\,cm^{-2}}$, and a redshift of $z = 2.151$. This yields $E_{\mathrm{iso}} = (5.48 \pm 0.22) \times 10^{52}$~erg (see Table~\ref{200613A_prop}). The rest-frame peak energy ($E_{\mathrm{p,i}} = E_{\mathrm{p}} \times (1+z)$) obtained from the spectral fitting of the time-integrated spectrum (see Table~\ref{200613A_prop}) is $E_{\mathrm{p,i}} = 106.6 \pm 6.7~\mathrm{keV}$. Fig.~\ref{fig:amati} shows the location of GRB~250129A in the Amati plane, which lies near the boundary between the long and ultra-long GRB populations.

\subsection{\texorpdfstring{$E_{\rm p}$--$T_{90}$}{Ep--T90} Correlation}
\label{hardness}

Another key correlation used to determine the position of GRB~250129A, is the spectral peak energy of the $\nu F_{\nu}$ spectrum and the $T_{90}$ duration of long and short GRBs. This relation, derived from the same GRB sample, exhibits a clear bimodal distribution corresponding to the two classes. The likelihood of a burst belonging to either class can be statistically evaluated using a Gaussian Mixture Model \cite{2017ApJ...848L..14G}. Based on the measured values of $E_{\rm p}$ and $T_{90}$, it is evident that GRB~250129A lies in the outer boundary of long GRBs, consistent with its position in the Amati correlation plane.

\section{Summary and conclusion}
\label{ch4:summary}
We conducted a comprehensive temporal and spectral analysis of the peculiar prompt emission of GRB~250129A. This burst, detected at a redshift of $z = 2.151$, represents an intermediate-luminosity GRB. It was initially triggered by the \textit{Swift}-BAT instrument, with subsequent follow-up observations performed by \textit{Swift}-XRT and UVOT. The BAT light curve exhibits a complex multi-peaked profile characterized by two prominent emission episodes. The measured duration of $T_{90} = 262.25 \pm 23.71$~s, derived from the 15–350~keV BAT light curve, places GRB~250129A near the transition between the long and ultra-long GRB classes, with the main emission peak occurring at approximately $T_0 + 180$~s.

The time-integrated BAT spectrum (15–150~keV) was modeled using various built-in phenomenological functions in 3ML, including the Band function, BPL, CPL, and BB models, along with their possible combinations. Among these, the BPL model provided the statistically best representation of the data, based on the BIC and log-evidence values, indicating a predominantly non-thermal origin with a relatively low peak energy ($E_{\rm p}$). To explore spectral evolution, the light curve was sliced into time intervals using the Bayesian block algorithm, ensuring sufficient photon statistics in each bin. The spectra corresponding to the second emission episode were consistently well-fitted by the BPL model, whereas only a few bins from the first episode yielded acceptable fits with the same model. The temporal evolution of the low-energy photon index ($\alpha$) closely follows the flux variation, demonstrating a clear intensity-tracking behavior. Further modeling and a detailed analysis of the flux–$\alpha$ correlation are required to better constrain the emission physics. The $\alpha$ value remains under the synchrotron slow-cooling limit throughout the burst, confirming its synchrotron origin.

Both the Amati relation and the hardness--intensity correlation place GRB~250129A close to the \(3\sigma\) boundary, indicating that it exhibits properties consistent with long GRBs while lying toward the extreme end of the long–GRB distribution in these correlations. A comparative investigation of GRBs located near the boundary between the long and ultra-long classes may provide important insights into their radiation mechanisms and the prolonged activity of their central engines.

\end{document}